\documentclass[conference]{IEEEtran}

\ifCLASSOPTIONonecolumn
\usepackage{setspace}
\fi

\IEEEoverridecommandlockouts
\usepackage{cite}
\usepackage{amsmath,amssymb,amsfonts}
\usepackage{adjustbox}
\usepackage{graphicx}
\usepackage{booktabs}
\usepackage{multirow}
\usepackage{placeins}
\usepackage{textcomp}
\usepackage{color}
\usepackage{comment}
\usepackage{psfig}
\usepackage{multirow}
\usepackage{steinmetz} 
\usepackage{tabularx}
\usepackage[multiple]{footmisc} 
\newcommand{\mycomment}[1]{%
}%
\usepackage{environ}
\NewEnviron{nestedcomment}{\mycomment{\BODY}}

\usepackage{algorithm}
\usepackage{algpseudocode}

\usepackage[font=small,labelfont=bf]{caption}

\usepackage{etoolbox}
\makeatletter
\patchcmd{\@makecaption}
  {\scshape}
  {}
  {}
  {}
\makeatother

\def\urltilde{\kern -.15em\lower .7ex\hbox{\~{}}\kern .04em}

\begin{document}
\title{Comments on ``A Linear Time Algorithm for the Optimal Discrete IRS Beamforming"
\thanks{The authors are with the Center for Pervasive Communications and Computing (CPCC), Department of Electrical Engineering and Computer Science,
University of California, Irvine.}
\thanks{This work is partially supported by NSF grant 2030029.}
}

\ifCLASSOPTIONonecolumn
\author{\IEEEauthorblockN{Dogan Kutay Pekcan, 
}
\IEEEauthorblockA{\textrm{CPCC, Department of EECS} \\
\textrm{University of California, Irvine, CA, USA}\\
dpekcan@uci.edu\\[7mm]}
\and
\IEEEauthorblockN{ Ender Ayanoglu, {\em Fellow, IEEE}}\\
\IEEEauthorblockA{\textrm{CPCC, Department of EECS} \\
\textrm{University of California, Irvine, CA, USA}
 \\ayanoglu@uci.edu}}
\else
\author{\IEEEauthorblockN{Dogan Kutay Pekcan}
\IEEEauthorblockN{Ender Ayanoglu, {\em Fellow, IEEE}}
}
\fi

\maketitle

\begin{abstract}
The problem of optimizing discrete phases in a reconfigurable intelligent surface (RIS) to maximize
the received power at a user equipment is addressed.
Comments on \cite{b1} are provided. Updated necessary and sufficient conditions for its Lemma~1 are given.
Consequently, an updated Algorithm~1 is provided with full specification. Simulation results with improved
performance over the implementation of Algorithm~1 are provided. New versions of the algorithm are given
that are proven to achieve convergence in $N$ or fewer steps, where $N$ is the number of the elements in the
reconfigurable intelligent surface. This is in contrast with $KN$ or $2N$ number of steps on the average
specified for Algorithm~1 in \cite{b1}, where $K$ is the number of discrete phases. As a result, for a discrete-phase RIS,
the techniques presented here achieve the optimum received power in the smallest number of steps published in the literature.
\end{abstract}

\section{Introduction}\label{ch:1}
Reference \cite{b1} presented an algorithm to solve the problem of finding the values $\theta_1,
\theta_2, \ldots, \theta_N$ to maximize $| h_0 + \sum_{n=1}^N h_n e^{j\theta_n} |$ where $\theta_n
\in \Phi_K$ and $\Phi_K = \{\omega, 2\omega, \ldots, K\omega\}$ with $\omega=\frac{2\pi}{K}$ and
$j=\sqrt{-1}$. The
set $\Phi_K$ can equivalently be described as $\{ 0, \omega, 2\omega, \ldots , (K-1)\omega\}$. In
\cite{b1}, the values $h_n\in \mathbb{C}$, $n=1,2,\ldots,N$ are the channel coefficients and
$\theta_n$ are the phase values added to the corresponding $h_n$ by an intelligent reflective
surface (IRS), also known as reconfigurable intelligent surface (RIS).

This paper is an extended version of \cite{PA23}.

\section{Two Statements from \cite{b1}}
Towards achieving its goal, \cite{b1} introduced the following lemma.

{\em Lemma 1:\/} For an optimal solution $(\theta_1^*, \ldots, \theta_n^*)$ to problem (8),
each $\theta_n^*$ must satisfy
\begin{equation}\tag{11}
\theta_n^* = \arg \min_{\theta_n\in \Phi_K} |(\theta_n +\alpha_n - \phase{\mu}) \;{\rm mod}\; 2\pi|
\label{eqn:eqn11}
\end{equation}
where $\phase{\mu}$ stands for the phase of $\mu$ in (10)\footnote{To prevent confusion,
we will use the same equation numbers (7)--(13) in \cite{b1}. Our own equation numbers,
not available in \cite{b1}, will begin at (19) and will be incremented from that number
on. Similarly, we will introduce Lemma~2 and Algorithm~2 in lieu of Lemma~1 and Algorithm~1
in \cite{b1}. Note that a lemma or an algorithm with
number 2 does not exist in \cite{b1}.}\footnote{In this paper, we define the ${\rm mod}$
function (the modulus function or the modulo operation) $x\; {\rm mod}\;y$ as the remainder after the
dividend $x>0$ is divided by the divisor $y>0$. We write it as $x\;{\rm mod}\; y$, $x\; (mod\; y)$, or
${\rm mod}\; (x, y)$. For $x<0$ and $y>0,$ we use the convention that the remainder should always
be the smallest such nonnegative number.}.

In \cite{b1}, problem (8) is defined as
\begin{align}
\tag{8a}\underset{\mbox{\boldmath$\theta$}}{\rm maximize\ } & f({\mbox{\boldmath$\theta$}})\\
\tag{8b}{\rm subject\ to\ } & \theta_n\in \Phi_K\quad {\rm for}\quad n=1, 2, \ldots, N
\end{align}
where
\begin{equation*}\tag{7b}
f({\mbox{\boldmath$\theta$}}) = \frac{1}{\beta_0^2}\bigg|\beta_0e^{j\alpha_0}+\sum_{n=1}^N \beta_n
e^{j(\alpha_n + \theta_n)}\bigg|^2,
\end{equation*}
$h_n = \beta_ne^{j\alpha_n}$ for $n = 0, 1, \ldots, N$,
and ${\mbox{\boldmath$\theta$}} = (\theta_1, \theta_2, \ldots, \theta_N)$. Also, $g$ is defined as
\begin{equation}\tag{9}
g = h_0 + \sum_{n=1}^N h_n e^{j\theta_n^*}
\end{equation} and $\mu$ as
\begin{equation}\tag{10}
\mu = \frac{g}{|g|}.
\end{equation}

\setcounter{equation}{18}
Lemma 1 does not hold. This can be seen by numerical examples. We give one such
example in Table~\ref{tbl:compare}. In this table, we look at the simple case of $K=2$, $N=2$.
According to Lemma 1 in \cite{b1}, the condition in (11) should satisfy (8) for this simple
case. We draw values of $h_n$ according to the first paragraph of Sec.~IV in \cite{b1}. We
list these values in rows 2--4 of Table~\ref{tbl:compare}. We define
\begin{equation}
g_0(\theta_1, \theta_2) = h_0 + \sum_{n=1}^2 h_n e^{j\theta_n}
\end{equation}
and list the values of $g_0(\theta_1,\theta_2)$ for all possible $\theta_1, \theta_2 \in \{0, \pi\}$.
There are four such values and they are listed in rows 5--8 of Table~\ref{tbl:compare}. The
set of values for $\theta_1$ and $\theta_2$ that maximize $|g_0|$, or equivalently, that
achieve $g$ in (9), are $\theta_1 = \theta_2 = \pi$ as in row 8 of Table~\ref{tbl:compare}.
Note that this operation results in $\phase{\mu}=2.3719$ radians as shown in column~5 of
row~8 of Table~\ref{tbl:compare}.

\begin{table*}[!t]
\begin{center}
\begin{tabular}{|c|c|c|c|c|}
\hline
 &${\rm Re}[\cdot]$&${\rm Im}[\cdot]$&$| \cdot |$&$\phase{\cdot}\in [0, 2\pi )$ (rad.)\\
 \hline
 $h_0$& $-2.8267\times 10^{-7}$ & $2.7376\times 10^{-7}$ & $3.9350\times 10^{-7}$ & $2.3722$\\
 $h_1$& $1.0958\times 10^{-10}$ & $-1.0501\times 10^{-11}$ & $1.1008\times 10^{-10}$ & $6.1876$\\
 $h_2$& $-1.2238\times 10^{-11}$ & $-2.6605\times 10^{-11}$ & $2.6634\times 10^{-10}$ & $4.6664$\\
 \hline
 $g_0(\theta_1=0,\theta_2=0)$&$-2.8257\times 10^{-7}$&$2.7348\times 10^{-7}$&$3.9324\times 10^{-7}$&$2.3725$\\
 $g_0(\theta_1=0,\theta_2=\pi)$&$-2.8255\times 10^{-7}$&$2.7401\times 10^{-7}$&$3.9359\times 10^{-7}$&$2.3715$\\
 $g_0(\theta_1=\pi,\theta_2=0)$&$-2.8279\times 10^{-7}$&$2.7350\times 10^{-7}$&$3.9341\times 10^{-7}$&$2.3729$\\
 $g_0(\theta_1=\pi,\theta_2=\pi)$&$-2.8277\times 10^{-7}$&$2.7403\times 10^{-7}$&${\bf 3.9377}\times {\bf 10^{-7}}$&$\bf 2.3719$\\
 \hline
\end{tabular}
\caption{Sample calculation for attempting to find optimum $\theta_1^*, \theta_2^*,\ldots,\theta_N^*$ to maximize $|g_0|$ where $g_0 (\theta_1, \theta_2, \ldots, \theta_N) = h_0 + \sum_{n=1}^N h_n e^{j\theta_n}$ with $\theta_n\in\Phi_K = \{0,\frac{2\pi}{K},\ldots,(K-1)\frac{2\pi}{K}\}$, $n=1, 2, \ldots, N$, for $K=2$ and $N=2$. Channel coefficients $h_n$, $n=0,1,2$ are calculated using the technique described in \cite{b1}.
Rows 5--8 present all values of $g_0$ with all combinations of $\theta_1, \theta_2 \in \Phi_2$, showing that $|g| = \max |g_0 (\theta_1, \theta_2) |$ is achieved with $\theta_1^* =\theta_2^* = \pi$.
}
\label{tbl:compare}
\end{center}
\end{table*}
\begin{table}[!t]
\begin{center}
\begin{tabular}{|c|c|}
\hline
$(\theta_1=0)+\alpha_1-\phase{\mu}$&{3.8158}\\
${\rm mod} ((\theta_1=0)+\alpha_1-\phase{\mu},2\pi)$&{3.8158}\\
$(\theta_1=\pi)+\alpha_1-\phase{\mu}$&{6.9574}\\
${\rm mod} ((\theta_1=\pi)+\alpha_1-\phase{\mu},2\pi)$&{\bf 0.67417}\\[0.7mm]
 \hline
$(\theta_2=0)+\alpha_2-\phase{\mu}$&{2.2945}\\
${\rm mod} ((\theta_2=0)+\alpha_2-\phase{\mu},2\pi)$&{\bf 2.2945}\\
$(\theta_2=\pi)+\alpha_2-\phase{\mu}$&{5.4361}\\
${\rm mod} ((\theta_2=\pi)+\alpha_2-\phase{\mu},2\pi)$&{5.4361}\\[0.7mm]
 \hline
$\cos((\theta_1=0)+\alpha_1-\phase{\mu})$&{-0.7812}\\
$\cos((\theta_1=\pi)+\alpha_1-\phase{\mu})$&{\bf 0.7812}\\[0.7mm]
 \hline
$\cos((\theta_2=0)+\alpha_2-\phase{\mu})$&{-0.6672}\\
$\cos((\theta_2=\pi)+\alpha_2-\phase{\mu})$&{\bf 0.6672}\\[0.7mm]
\hline
\end{tabular}
\caption{Continuation of the sample calculation for attempting to find optimum $\theta_1^*, \theta_2^*,\ldots,\theta_N^*$ to maximize $|g_0|$. Rows 1--8 present the calculation of $\min_{\theta_n\in \Phi_K} \;{\rm mod}\; (\theta_n + \alpha_n - \phase{\mu}, 2\pi)$ for $n=1,2,\ldots,N$, as specified in \cite{b1} to attempt to find the optimum values of $\theta_n$. This calculation results in values $\theta_1=0$ and $\theta_2=\pi$, which are not $\theta_1^*, \theta_2^*$. Rows 9-12 present the calculation of $\max_{\theta_n\in \Phi_K} \cos(\theta_n + \alpha_n - \phase{\mu})$ to find $\theta_1^*, \theta_2^*, \ldots, \theta_N^*$ as discussed in this comment. This technique finds the optimum values of $\theta_n,$ $n=1,2,\ldots, N$.
}
\label{tbl:compare2}
\end{center}
\end{table}
At this point, we would like to emphasize that \cite{b1} uses a particular convention for the
phases of complex numbers. They are defined to be in $[0, 2\pi)$, see the text that follows
(2) in \cite{b1}. We use the same convention in generating Table~\ref{tbl:compare}, see its
column 5, as well as in generating Table~\ref{tbl:compare2}. With this convention, we list
 $\theta_n + \alpha_n -\phase{\mu}$ and
$(\theta_n + \alpha_n - \phase{\mu}) \;{\rm mod}\; 2\pi$ for possibilities of $\theta_n = 0$ and
$\theta_n=\pi$ and
$n=1,2$ in rows 1--8 of Table~\ref{tbl:compare2}.\footnote{Note that absolute value
signs in (11) are not needed since the argument of the minimum operation
in (11) is in $[0, 2\pi)$.} It can be seen from rows~1--4 of Table~\ref{tbl:compare2}
that the method results in $\theta_1 = \pi$ as the potential $\theta_1^*$, which we
know from the discussion in the previous paragraph to be correct. When we carry out
the calculation $(\theta_2 + \alpha_2 - \phase{\mu}) \; {\rm mod}\; 2\pi$ in rows 5--8 of
Table~\ref{tbl:compare2}, we find that the method suggests $\theta_2=0$ should be
$\theta_2^*$. However, we know from the exhaustive search in rows 5--8 of
Table~\ref{tbl:compare} that $\theta_2^*=\pi$. Thus, Lemma~1 is not correct.

It is possible to come up with a correct lemma similar to Lemma~1. We
specify this lemma below.

{\em Lemma 2:\/} For an optimal solution $(\theta_1^*, \theta_2^*, \ldots,
\theta_n^*)$, it is necessary and sufficient that each $\theta_n^*$ satisfy
\begin{equation}
\theta_n^* = \arg \max_{\theta_n\in \Phi_K} \cos(\theta_n + \alpha_n -\phase{\mu})
\label{eqn:lemma2}
\end{equation}
where $\phase{\mu}$ stands for the phase of $\mu$ in (10).

{\em Proof:\/} We can rewrite (9) as
\begin{align}
|g| =& \ \beta_0 e^{j(\alpha_0-\phase{\mu})} + \sum_{n=1}^N \beta_n e^{j(\alpha_n+\theta_n-\phase{\mu})} \\
 = & \ \beta_0 \cos(\alpha_0 - \phase{\mu}) + j \beta_0 \sin (\alpha_0-\phase{\mu}) \nonumber\\
 & + \sum_{n=1}^N \beta_n \cos(\theta_n + \alpha_n - \phase{\mu}) \nonumber\\
 & + j \sum_{n=1}^N \beta_n \sin(\theta_n + \alpha_n - \phase{\mu}).
 \label{eqn:absg}
\end{align}
Because $|g|$ is real-valued, the second and fourth terms in (\ref{eqn:absg}) sum to zero, and
\begin{equation}
|g| = \beta_0 \cos(\alpha_0 - \phase{\mu}) + \sum_{n=1}^N \beta_n \cos(\theta_n + \alpha_n - \phase{\mu})
\end{equation}
from which (\ref{eqn:lemma2}) follows as a necessary and sufficient condition for Lemma~2 to hold.
\hfill$\blacksquare$

Rows~9--12 of Table~\ref{tbl:compare2} illustrate that this method finds $\theta_1^*$ and $\theta_2^*$.
More extensive calculations can be carried out to show that an exhaustive search as in rows 5--8 of
Table~\ref{tbl:compare} confirms that Lemma~2 holds for a wide set of $K$ and $N$ values as well as
a wide set of channel coefficients $h_0, h_1, \ldots, h_N$.

In addition to (\ref{eqn:eqn11}) not being the correct measure, the proof of Lemma~1 is not
accurate. This problem appears not only in \cite{b1}, but also in its versions \cite{b11,b12,b13,b14,b15} 
on arxiv.org, some
of which appeared after \cite{b1}. Two of these references \cite{b14,b15} have different measures 
than (\ref{eqn:eqn11}), but their proofs have the same issue with that in \cite{b1}. For a 
discussion on this subject, see Appendix~\ref{sec:app-prooflemma}.

Reference \cite{b1} attempts to decide a range of $\mu$ for which $\theta_n^* = k\omega$ must hold,
making use of Lemma~1. Towards that end, it first defines a sequence of complex numbers with respect to
each $n=1,2,\ldots,N$ as
\begin{equation}\tag{12}
s_{nk} = e^{j(\alpha_n+(k-0.5)\omega)},\ {\rm for}\ k=1,2,\ldots,K.
\label{eqn:eqn12}
\end{equation}
Then, \cite{b1} defines, for any two points $a$ and $b$ on the unit circle $C$, ${\rm arc}(a:b)$
to be the unit circular arc with $a$ as the initial end and $b$ as the terminal end in the counterclockwise
direction; in particular, it defines ${\rm arc}(a:b)$ as an open arc with the two endpoints $a$ and $b$
excluded. With this definition, \cite{b1} states the following proposition follows from Lemma 1.

{\em Proposition 1:\/} A sufficient condition for $\theta_n^*=k\omega$ is
\begin{equation}\tag{13}
\mu \in {\rm arc} (s_{nk}:s_{n,k+1}).
\end{equation}
Reference \cite{b1} states that ``letting $\theta_n = k\omega$ is guaranteed to minimize the gap
$|(\theta_n+\alpha_n-\phase{\mu}) \;{\rm mod}\; 2\pi |$ whenever $\mu$ lies in its associated arc, and thus
$k\omega$ must be optimal according to Lemma 1.''

Now, let $K=2$ and thus $\omega=\frac{2\pi}{K}=\pi$, and the two possibilities for $\theta$ are
$\theta^1=\pi$ and $\theta^2 = 2\pi$, or equivalently $\theta^2 = 0$. According to (12), we have
\begin{equation}
s_{n1}=e^{j(\alpha_n+\frac{\pi}{2})},\quad s_{n2}=e^{j(\alpha_n+\frac{3\pi}{2})}.
\end{equation}
According to Proposition~1, if $\mu\in {\rm arc}(s_{n1}:s_{n2})$ then $\theta_n^*=\omega=\pi$
should hold. Assume $\mu$ is in ${\rm arc}(s_{n1},s_{n2})$. Then, it can be observed that
$\alpha_n-\phase{\mu}\in (\frac{\pi}{2},\frac{3\pi}{2})$, paying
attention to the change of order due to the subtraction of $\phase{\mu}$. In particular, let
$\mu$ be
such that $\alpha_n-\phase{\mu}\in (\frac{\pi}{2},\pi)$. When this is the case, note that $(\theta^1
+\alpha_n - \phase{\mu})\in (\frac{3\pi}{2}, 2\pi)$ while $(\theta^2+\alpha_n-\phase{\mu})\in
(\frac{\pi}{2},\pi)$. Thus, $|(\theta^2 + \alpha_n - \phase{\mu}) \;{\rm mod}\; 2\pi| <
|(\theta^1 + \alpha_n - \phase{\mu}) \;{\rm mod}\; 2\pi|$, and according to Lemma~1,
$\theta_n^*=\theta^2=0$, in contradiction with Proposition~1.
On the other hand, Proposition~1 is compatible with Lemma~2. To see this, assume
$\mu$ satisfies (12). Then,
\begin{equation}
\phase{\mu}\in \Big(\alpha_n+\Big(k-\frac{1}{2}\Big)\omega,\alpha_n+\Big(k+\frac{1}{2}\Big)\omega\Big).
\end{equation}
Since $\omega=\frac{2\pi}{K}$,
\begin{equation}
\alpha_n-\phase{\mu}\in \Big((-2k-1)\frac{\pi}{K}, (-2k+1)\frac{\pi}{K}\Big)
\end{equation}
considering the reversal of order due to the substraction of $\phase{\mu}$.
Now, let $\theta_n = k\omega = 2k \frac{\pi}{K}$. Then
\begin{equation}
\theta_n + \alpha_n - \phase{\mu} \in \Big(-\frac{\pi}{K}, \frac{\pi}{K}\Big)
\end{equation}
and thus $\cos(\theta_n + \alpha_n-\phase{\mu})$ is the largest among
all other possibilities for $\theta_n$ because the slice $(-\frac{\pi}{K}, \frac{\pi}{K})$ corresponds
to the largest values of the cosine function among all slices corresponding to different values of
$\theta_k \in \Phi_K$ for $n=1,2,\ldots,K$.

\section{A New Algorithm}
\setcounter{algorithm}{1}
\begin{algorithm}[!t]
\caption{Update for Algorithm~1 \cite{b1}}\label{alg:alg2}
\begin{algorithmic}[1]
\State {\bf Initialization:} Compute 
$s_{nk}=e^{j(\alpha_n + (k -0.5)\omega)}$ for $n=1,2,\ldots,N$ and
$k=1,2,\ldots,K$.
\State Eliminate duplicates among $s_{nk}$ and sort to get $e^{j\lambda_l}$
such that $0\le \lambda_1 < \lambda_2 < \cdots < \lambda_L < 2\pi.$
\State Let, for $l = 1,2,\ldots,L,$ ${\cal N} (\lambda_l) =
\{ n | \phase{s_{nk}} = \lambda_l \}.$
\State 
Set $\phase{\mu} = 0$. For $n=1,2,\ldots,N$, calculate
$\theta_n = \arg\max_{\theta_n\in\Phi_K} \cos(\theta_n + \alpha_n - \phase{\mu})$.
\State Set $g_0 = h_0 + \sum_{n=1}^N h_ne^{j\theta_n}$, ${\tt absgmax} = |g_0|$.
\For{$l = 1, 2, \ldots, L-1$}
\State For each $n\in{\cal N}(\lambda_l)$, let $(\theta_n + \omega \leftarrow \theta_n) \;{\rm mod}\;\Phi_K$.
\State Let
\[
g_l = g_{l-1} + \sum_{n\in {\cal N}(\lambda_l)} h_n \big(e^{j\theta_n} - e^{j(\theta_n - \omega) \;{\rm mod}\; \Phi_K}\big)
\]
\If{$|g_l| > {\tt absgmax}$}
\State Let ${\tt absgmax} = |g_l|$
\State Store $\theta_n$ for $n=1,2,\ldots,N$
\EndIf
\EndFor
\State Read out $\theta_n^*$ as the stored $\theta_n$, $n=1,2,\ldots,N$.
\end{algorithmic}
\end{algorithm}

We now specify Algorithm~2 to replace Algorithm~1 in \cite{b1}. In
doing so, not only do we incorporate Lemma~2 instead of Lemma~1 but
also we eliminate the many uncertainties present in Algorithm~1 of
\cite{b1}.

In Algorithm~\ref{alg:alg2} we define $(\theta \pm \omega)\,{\rm mod}\,\Phi_K$ as follows. First note that the two
sets $\{0,\omega,2\omega,\dots,(K-1)\omega\}$ and $\{\omega,2\omega,3\omega,\ldots,K\omega\}$ have the same
members since $\omega = 2\pi/K$. Considering the first set above, when $\theta=k\omega$, $k=0,1,2,\ldots,K-1$,
\begin{equation*}
(\theta + \omega)\,{\rm mod}\,\Phi_K \triangleq \left\{\begin{array}{ll}(k+1)\omega&{\rm if\ }k=0,1,\ldots,K-2,\\
0&{\rm if\ }k=K-1,\end{array}\right.
\end{equation*}
and
\begin{equation*}
(\theta - \omega)\,{\rm mod}\,\Phi_K \triangleq \left\{\begin{array}{ll}(K-1)\omega&{\rm if\ }k=0,\\
(k-1)\omega&{\rm if\ }k=1,2,\ldots,K-1.\end{array}\right.
\end{equation*}
Or, it can be expressed compactly as
\begin{equation}
(\theta \pm \omega)\,{\rm mod}\,\Phi_K \triangleq ((k \pm 1)\,{\rm mod}\,K)\,\omega .
\end{equation}

In Appendix~\ref{sec:appSimpleUpdate}, we discuss an alternative technique to initialize Algorithm~2. In addition to 
Algorithm~2, we will use this technique in initializing Algorithm~3 and Algorithm~4 in the sequel.

\section{Results and Remarks}
Because its description is based on Lemma~1, which does not provide an equivalency condition
for finding $\theta_1^*, \theta_2^*, \ldots, \theta_N^*$, the performance of Algorithm~1 will
in general not achieve the optimum result for SNR Boost \cite{b1}.

\begin{figure}[!t]
\centering
\includegraphics[width=0.45\textwidth]{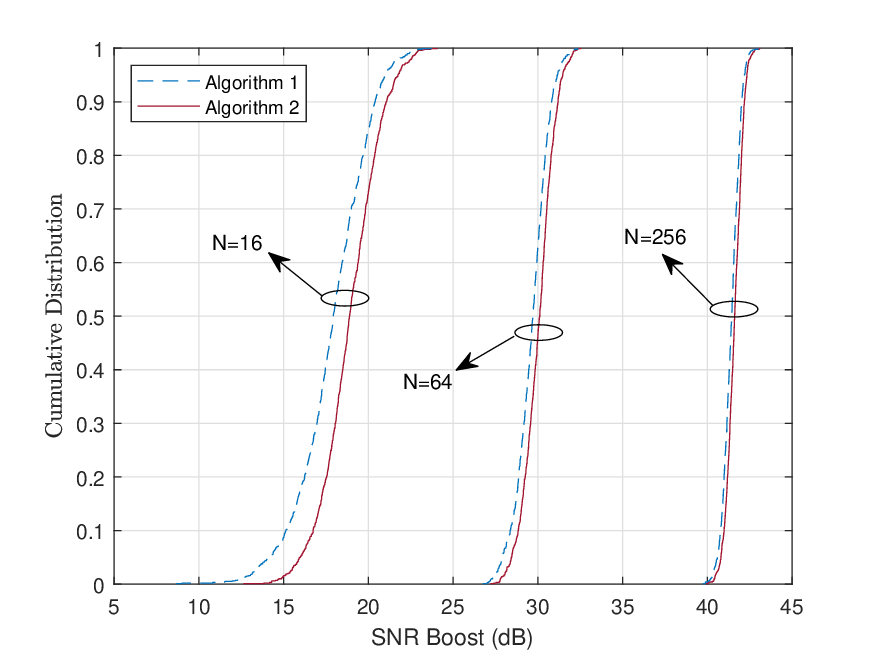}
\caption{CDF plots for SNR Boost with Algorithm 1 \cite{b1} and Algorithm 2, $K=2$.}
 \label{fig:SNRBoostK2}
 \end{figure}
We have implemented Algorithm~1 to the best of our interpretation. We have also implemented
Algorithm~2.
We present the CDF results for SNR Boost \cite{b1} in Fig.~\ref{fig:SNRBoostK2} for $K=2$
and $N =16$, $64$, and $256$, using the average of 1,000 realizations of the channel.
Clearly,
Algorithm~1 is not optimal. Algorithm~2 performs better than Algorithm~1 although the
gains decrease with $N$. Plots for $K=4$ show smaller gains as compared to $K=2$, but
still, Algorithm~2 always performs better than Algorithm~1 for the same $K$ and $N$.

We note that it is possible to convert the maximization of $\cos(\theta_n+\alpha_n-\phase{\mu})$
to the minimization of a simple expression. For example, minimization of $f_1(x)=
\pi - | (x \;{\rm mod}\; 2\pi ) - \pi |$ is the same as maximization of $\cos(x)$ within
the context of Lemma~2. However, this is different than minimization of $|x \;{\rm mod}\; 2\pi|$
proposed in Lemma~1 of \cite{b1}. The reason can be
seen by plotting these functions against $x$. While $f_1(x)$ and $\cos(x)$, in
addition to being periodic with period $2\pi$, have even symmetry around odd
multiples of $\pi$, $| x \; {\rm mod}\; 2\pi |$ (or equivalently, $(x \; {\rm mod}\; 2\pi)$) does not
have this symmetry.

\section{Algorithm Convergence: Towards Two New Algorithms}
We will now prove that the algorithm takes $N$ or fewer steps to converge,
as opposed to the statement in \cite{b1} that it takes $KN$ or $2N$ steps on
average. 
Towards this end, we first make the following statement.

{\em Claim 1:\/} As in (\ref{eqn:eqn12}), set $s_{nk}=e^{j(\alpha_n+(k-\frac{1}{2})\frac{2\pi}{K})}$,
$n=1,2,\ldots, L \le NK,$ $k=1,2,\ldots,K$, $\alpha_n\in [0,2\pi ).$ Let $\lambda_l = \phase{s_{nk}}$ such
that $0\le \lambda_1 < \lambda_2 < \cdots < \lambda_L <2\pi$. Let ${\cal N}(\lambda_l)=\{n|\lambda_l = \phase{s_{nk}}\}$.
Assuming for now that $|{\cal N}(\lambda_l)| = 1$, $l=1,2\ldots,L$, which we will relax in the sequel,
we claim that ${\cal N}(\lambda_{l'}) = {\cal N}(\lambda_{l'+N})$ for $l'=1,2,\ldots,N(K-1).$

To prove {\em Claim 1,\/} we will first introduce {\em Claim~2\/} and prove it.

{\em Claim 2:\/} Without loss of generality, we can assume that $\alpha_n < \frac{2\pi}{K},$ $n=1,2,\ldots,L=NK$.

{\em Proof of Claim 2:\/} Suppose that for some $n$, we have $\frac{2\pi}{K}\cdot m \le \alpha_n < \frac{2\pi}{K}\cdot (m+1),$
$m=1, 2, \ldots, K-1$. Let $\beta_n \triangleq \alpha_n - m\cdot \frac{2\pi}{K},$ so that $\beta_n< \frac{2\pi}{K}$.
We will write below each phase value in (\ref{eqn:eqn12}), $\alpha_n + \frac{(2k-1)\pi}{K}$, for $k=1,2,\ldots,K$ (note that
the $({\rm mod}\; 2\pi)$ notation below applies to both sides of the equation).
\begin{itemize}
\item $k=1\hspace{-1mm}:$\\ $\alpha_n+\frac{\pi}{K} = \beta_n+\frac{(2m+1)\pi}{K} \ ({\rm mod}\; 2\pi)$\\
\hspace{15mm}where $\Big(\frac{(2m+1)\pi}{K}\Big)_{m=1}^{K-1} = \Big\{\frac{3\pi}{K},\frac{5\pi}{K},\ldots,\frac{(2K-1)\pi}{K}\Big\},$
\item $k=2\hspace{-1mm}:$\\ $\alpha_n+\frac{3\pi}{K} = \beta_n + \frac{(2m+3)\pi}{K}\ ({\rm mod}\; 2\pi)\\ =\left\{\begin{tabular}{ll}
$\beta_n +\frac{(2m+3)\pi}{K}\ ({\rm mod}\; 2\pi),$&$m \le K-2$\\
$\beta_n +\frac{\pi}{K}\ ({\rm mod}\; 2\pi),$&$m = K-1$
\end{tabular}\right.$\\
\hspace{15mm}where $\Big(\frac{(2m+3)\pi}{K}\Big)_{m=1}^{K-2} = \Big\{\frac{5\pi}{K},\frac{7\pi}{K},\ldots,\frac{(2K-1)\pi}{K}\Big\},$\\
$\vdots$
\item $k=K\hspace{-1mm}:$\\ $\alpha_n+\frac{(2K-1)\pi}{K} = \beta_n +\frac{(2m+2K-1)\pi}{K}\ ({\rm mod}\; 2\pi)\\
\mbox{\hspace{20.8mm}}=\beta_n+\frac{(2m-1)\pi}{K}\ ({\rm mod}\; 2\pi)$ \\
\hspace{15mm}where $\Big(\frac{(2m-1)\pi}{K}\Big)_{m=1}^{K-1} = \Big\{\frac{\pi}{K},\frac{3\pi}{K},\ldots,\frac{(2K-3)\pi}{K}\Big\}.$
\end{itemize}
Thus, if there is an $\alpha_n\ge \frac{2\pi}{K}$ to generate $K$ phase values, there is always a $\beta_n$, $\beta_n< \frac{2\pi}{K}$
with which one can generate the same $K$ phase values in a similar fashion. Therefore, in order to
prove Claim 1, one can work with the assumption that $\alpha_n<\frac{2\pi}{K},$ for $n=1,2,\ldots, L$.\hfill$\blacksquare$

{\em Proof of Claim 1:\/} 
Assuming $0\le \alpha_1 < \alpha_2 < \cdots < \alpha_L<\frac{2\pi}{K},$ $n=1,2,\ldots,L$ without loss of generality,
we will now show that
${\cal N}(\lambda_{l'}) = {\cal N}(\lambda_{l'+N})$ for $l'=1,2,\ldots,N(K-1).$ For this, there are $N+1$ cases to consider.

{\em Case 0:\/} In this case, we write all possible values of $\phase{s_{nk}}$ as follows.
\begin{flushleft}
$n=1\hspace{-1mm}:$
\end{flushleft}
\vspace{-5mm}
\begin{equation}
\phase{s_{1k}}\in \textstyle\left\{ \alpha_1+ \frac{\pi}{K}, \alpha_1 +\frac{3\pi}{K}, \ldots , \alpha_1 + \frac{(2K-1)\pi}{K}\right\},
\label{eqn:eqn28}
\end{equation}
\begin{flushleft}
$n=2\hspace{-1mm}:$
\end{flushleft}
\vspace{-5mm}
\begin{equation}
\phase{s_{2k}}\in \textstyle\left\{ \alpha_2+ \frac{\pi}{K}, \alpha_2 +\frac{3\pi}{K}, \ldots , \alpha_2 + \frac{(2K-1)\pi}{K}\right\},
\label{eqn:eqn29}
\end{equation}
\hspace{7mm}$\vdots$
\begin{flushleft}
$n=N\hspace{-1mm}:$
\end{flushleft}
\vspace{-5mm}
\begin{equation}
\hspace{4mm}\phase{s_{Nk}}\in \textstyle\left\{ \alpha_N+ \frac{\pi}{K}, \alpha_N +\frac{3\pi}{K}, \ldots , \alpha_N + \frac{(2K-1)\pi}{K}\right\}.
\label{eqn:eqn30}
\end{equation}
Sorting (\ref{eqn:eqn28})--(\ref{eqn:eqn30}), we have
\begin{equation}
\begin{aligned}
\hspace{2mm} & \textstyle \alpha_1 + \frac{\pi}{K} < \alpha_2 + \frac{\pi}{K} < \cdots < \alpha_N + \frac{\pi}{K} \\
& < \textstyle \alpha_1 + \frac{3\pi}{K} < \alpha_2 + \frac{3\pi}{K} < \cdots < \alpha_N + \frac{3\pi}{K} \\
& \hspace{2mm}\vdots\\
& < \textstyle \alpha_1 + \frac{(2K-1)\pi}{K} < \alpha_2 + \frac{(2K-1)\pi}{K} < \cdots < \alpha_N + \frac{(2K-1)\pi}{K} .\nonumber
\end{aligned}
\end{equation}
Thus,
\begin{equation*}
\big({\cal N}(\lambda_l)\big)_{l=1}^{L=NK} = \Big\{ \underbrace{1, 2, \ldots, N}_{1}, \underbrace{1, 2, \ldots, N}_{2}, \ldots,
\underbrace{1, 2, \ldots, N}_{K}\Big\}.
\end{equation*}
Therefore, for Case 0 and for $l'=1,2,\ldots,N(K-1),$ we have ${\cal N}(\lambda_{l'}) = {\cal N}(\lambda_{l'+N})$.

There are $N$ remaining cases. We will discuss these cases as {\em Case $i$\/} where $i = 1, 2 , \ldots , N$.

{\em Case $i, (i=1,2,\ldots, N)\!\! :$\/} In {\em Case $i$,\/} we have $i$ occurrences of $\alpha_n > \frac{\pi}{K}$ as
follows.
\begin{equation}
\begin{aligned}
0 & < \alpha_1 < \alpha_2 < \cdots < \alpha_{N-i} < \frac{\pi}{K} \\
& < \alpha_{N-i+1} < \cdots < \alpha_N < \frac{2\pi}{K}.
\label{eqn:eqn34}
\end{aligned}
\end{equation}
We write all possible values of $\phase{s_{nk}}$ as follows.
\begin{flushleft}
$n=1\hspace{-1mm}:$
\end{flushleft}
\vspace{-5mm}
\begin{equation}
\begin{aligned}
\hspace{-2mm}\phase{s_{1k}}\in \textstyle\Big\{ \alpha_1+ & \frac{\pi}{K}, \alpha_1 +\frac{3\pi}{K}, \ldots , \\
& \alpha_1 + \frac{(2K-3)\pi}{K}, \alpha_1 + \frac{(2K-1)\pi}{K}\Big\},
\end{aligned}
\label{eqn:eqn35}
\end{equation}
$\hspace{5mm}\vdots$
\begin{flushleft}
$n=N-i\hspace{-1mm}:$
\end{flushleft}
\vspace{-5mm}
\begin{equation}
\begin{aligned}
\phase{s_{(N-i)k}}\in \textstyle\Big\{ & \alpha_{N-i}+ \frac{\pi}{K}, \alpha_{N-i} +\frac{3\pi}{K}, \ldots , \\
 & \hspace{-10mm}\alpha_{N-i} + \frac{(2K-3)\pi}{K}, \alpha_{N-i} + \frac{(2K-1)\pi}{K}\Big\},
\end{aligned}
\label{eqn:eqn36}
\end{equation}
\begin{flushleft}
$n=N-i+1\hspace{-1mm}:$
\end{flushleft}
\vspace{-5mm}
\begin{equation}
\begin{aligned}
\phase{s_{(N-i+1)k}}\in \textstyle\Big\{ & \alpha_{N-i+1}+ \frac{\pi}{K}, \alpha_{N-i+1} +\frac{3\pi}{K}, \ldots , \\
 & \hspace{-10mm}\alpha_{N-i+1} + \frac{(2K-3)\pi}{K}, \alpha_{N-i+1} - \frac{\pi}{K}\Big\},
\end{aligned}
\label{eqn:eqn37}
\end{equation}
\begin{flushleft}
$n=N-i+2\hspace{-1mm}:$
\end{flushleft}
\vspace{-5mm}
\begin{equation}
\begin{aligned}
\phase{s_{(N-i+2)k}}\in \textstyle\Big\{ & \alpha_{N-i+2}+ \frac{\pi}{K}, \alpha_{N-i+2} +\frac{3\pi}{K}, \ldots , \\
 & \hspace{-10mm}\alpha_{N-i+2} + \frac{(2K-3)\pi}{K}, \alpha_{N-i+2} - \frac{\pi}{K}\Big\},
\end{aligned}
\label{eqn:eqn38}
\end{equation}
$\hspace{5mm}\vdots$
\begin{flushleft}
$n=N\hspace{-1mm}:$
\end{flushleft}
\vspace{-5mm}
\begin{equation}
\begin{aligned}
\hspace{-2mm}\phase{s_{Nk}}\in \Big\{ \alpha_N+ \frac{\pi}{K}, & \alpha_N +\frac{3\pi}{K}, \ldots , \\
 & \alpha_N + \frac{(2K-3)\pi}{K}, \alpha_N - \frac{\pi}{K}\Big\}.
\end{aligned}
\label{eqn:eqn39}
\end{equation}
Sorting (\ref{eqn:eqn35})--(\ref{eqn:eqn39}), we have
\begin{equation*}
\alpha_{N-i+1} - \frac{\pi}{K} < \cdots < \alpha_N - \frac{\pi}{K} \hspace{27mm}
\end{equation*}
\begin{equation*}
\hspace{30mm}< \alpha_1 +\frac{\pi}{K} < \cdots < \alpha_{N-i} + \frac{\pi}{K}
\end{equation*}
\begin{equation*}
< \alpha_{N-i+1} + \frac{\pi}{K} < \cdots < \alpha_N + \frac{\pi}{K} \hspace{30mm}
\end{equation*}
\begin{equation*}
\hspace{30mm}< \alpha_1 +\frac{3\pi}{K} < \cdots < \alpha_{N-i} + \frac{3\pi}{K}
\end{equation*}
\begin{equation*}
< \alpha_{N-i+1} + \frac{3\pi}{K} < \cdots < \alpha_N + \frac{3\pi}{K} \hspace{30mm}
\end{equation*}
\begin{equation*}
\hspace{30mm}< \alpha_1 +\frac{5\pi}{K} < \cdots < \alpha_{N-i} + \frac{5\pi}{K}
\end{equation*}
\begin{equation*}
\hspace{-50mm}\vdots
\end{equation*}
\begin{equation*}
< \alpha_{N-i+1} + \frac{(2K-3)\pi}{K} < \cdots < \alpha_N + \frac{(2K-3)\pi}{K} \hspace{1mm}
\end{equation*}
\begin{equation*}
\hspace{15mm}< \alpha_1 +\frac{(2K-1)\pi}{K} < \cdots < \alpha_{N-i} + \frac{(2K-1)\pi}{K}
\end{equation*}
Thus,
\begin{align}
\big({\cal N}(\lambda_l)\big) & _{l=1}^{L=NK} = \nonumber\\
\Big\{ & \underbrace{N-i+1,\ldots,N,1, 2, \ldots, N-i}_{1},\nonumber\\
& \underbrace{N-i+1,\ldots,N,1, 2, \ldots, N-i}_{2}, \label{eqn:eqn387}\\
& \hspace{24.5mm}\vdots \nonumber\\
& \underbrace{N-i+1,\ldots,N,1, 2, \ldots, N-i}_{K}\Big\}\nonumber
\end{align}
for $i=1, 2 , \ldots, N$.
Therefore, for Case $i$, $i=1,2,\ldots,N$, and for $l'=1,2,\ldots,N(K-1),$ we have ${\cal N}(\lambda_{l'}) = {\cal N}(\lambda_{l'+N})$.
With this, Claim 1 is proved.
\hfill$\blacksquare$

\section{$N$ Steps Suffice When $|{\cal N}(\lambda_l)|=1$ for All $l$}\label{sec:nsteps}
Given $|{\cal N}(\lambda_l)|=1$ and
\begin{equation}
{\cal N}(\lambda_{l'}) = {\cal N}(\lambda_{l'+N}),\quad l'=1,2,\dots,N(K-1),
\label{eqn:eqn371}
\end{equation}
we want to show that $N$ steps will suffice for convergence.
Now, consider the main problem of maximizing $\left|h_0 + \sum_{n=1}^N h_n e^{j\theta_n}\right|$, where it is clear that our discrete
phase shift selections can only tune the second term in the absolute value. Let
\begin{equation}
g_c \triangleq \sum_{n=1}^N h_ne^{j\theta_n} = \sum_{n=1}^N \beta_n e^{j(\alpha_n+\theta_n)} .
\label{eqn:eqn395}
\end{equation}
In each step of the Algorithm~2, we define
\begin{equation}
g_{c,l} \triangleq g_l - h_0,\quad l=1,2,\ldots,L.
\label{eqn:eqn391}
\end{equation}
Note that, in (\ref{eqn:eqn391}), $h_0$, $g_l$, and $g_{c,l}$ are complex numbers, with $l$ being a generation index. We know that
whenever $\mu$ is anywhere in ${\rm arc}(s_{nk}:s_{n,k+1})$, $\theta_n$ does not change. The angle $\theta_n$
only changes when $\mu$ changes from one arc to another, i.e.,
\begin{equation}
\mu\in {\rm arc}{(e^{j\lambda_l}:e^{j\lambda_{l+1}})} \rightarrow \mu\in {\rm arc}(e^{j\lambda_{l+1}}:e^{j\lambda_{l+2}})
\label{eqn:eqn40}
\end{equation}
in which case $\theta_n$ must be updated as
\begin{equation}
\theta_n \rightarrow \theta_n + \omega, \quad n\in {\cal N}(\lambda_{l+1}).
\label{eqn:eqn41}
\end{equation}
With (\ref{eqn:eqn40})--(\ref{eqn:eqn41}), the naive approach in Algorithm~2 gathers all possibilities for 
$g_l$ in $NK$
steps by considering all possible arcs that $\mu$ can be in.
To show that $N$ steps will suffice, we want to point out the redundancy in those $NK$ steps.
Consider any consecutive $N$ steps in Algorithm~2.
In those steps, the phase shifts will be updated as $\theta_n \rightarrow \theta_n+\omega$
%
%
with $n\in ({\cal N}(\lambda_l))_{l=l'}^{l'+N}$, $l'=1,2,\ldots ,N(K-1)$. Since we have (\ref{eqn:eqn371}),
the following must hold
\begin{equation}
({\cal N}(\lambda_l))_{l=l'}^{l'+N} = \{1, 2, \ldots, N\} ,
\end{equation}
which says that after {\em any\/} $N$ consecutive steps in Algorithm~2, $\theta_n \rightarrow \theta_n + \omega,$
$n = 1, 2, \ldots, N$. To proceed further, we need an intermediate result, which we discuss below.

{\em Claim 3:\/} Let $g_c=\sum_{n=1}^N \beta_n e^{j(\theta_n + \alpha_n)}$ be the cascaded channel term in (\ref{eqn:eqn395}).
For any angle $\theta'$,
\begin{equation}
|g_c(\theta_1,\theta_2,\ldots,\theta_n)| = |g_c(\theta_1+\theta',\theta_2+\theta',\ldots,\theta_N+\theta')|.
\end{equation}

{\em Proof of Claim 3:\/} Let us write $| g_c |^2$ as follows
\begin{eqnarray}
\!\!|g_c|^2 \!\! &\!\!\! = \!\!\! & \left|\sum_{n=1}^N\beta_n e^{j(\theta_n+\alpha_n)}\right|^2 \nonumber\\
 & \!\!\! = \!\!\! & \sum_{k=1}^N \beta_k^2 \nonumber\\
 & \!\!\! \!\!\! & + \; 2 \sum_{k=2}^N\sum_{l=1}^{k-1} \beta_k\beta_l \cos((\theta_k + \alpha_k)-(\theta_l+\alpha_l))
 \label{eqn:eqn45}
\end{eqnarray}
If $\theta_n \rightarrow \theta_n + \theta',$ $n=1,2,\ldots,N$, then the angles $\theta'$ will cancel inside the cosine in
(\ref{eqn:eqn45}). This proves {\em Claim~3.\/}
$\hfill\blacksquare$

With {\em Claim~3, \/} we established
that if $\theta_n \rightarrow \theta_n + \omega$ with $n=1,2,\ldots,N$,
\begin{equation}
|g_{c,l'}| = |g_{c,l'+N}|,\quad l'=1,2,\ldots, N(K-1)
\label{eqn:eqn43}
\end{equation}
must be true. Therefore, among the $NK$ possibilities in Algorithm~2, there are only $N$ unique values of $|g_c|$.
Consequently, as the algorithm is tuning $g_c$ to maximize $|g_c+h_0|$, it is sufficient to consider $N$ arcs that are closest
to $h_0$.

The algorithm to implement when ${\cal N}(\lambda_l) = 1$ for $l=1,2,\ldots,L$  is given under Algorithm~\ref{alg:alg31}.
The initialization technique introduced in Appendix~\ref{sec:appSimpleUpdate} for Algorithm~2 is employed in Algorithm~3.
\begin{algorithm}[!t]
\caption{Simplified Algorithm 2 with $|{\cal N}(\lambda_l)| =1$ for all $l$\label{alg:alg31}}
\begin{algorithmic}[1]
\State {\bf Initialization:} Set $\phase{\mu}=\alpha_0 - \frac{\pi}{K}$ 
\State Compute $\varphi_n = (\alpha_n - \alpha_0)$ $\mathrm{mod}\, \frac{2\pi}{K}$, $n=1,2,\ldots,N$ 
\State Sort $\varphi_n$ such that $0 \le \varphi_{n_1} < \varphi_{n_2} < \cdots < \varphi_{n_N} <\frac{2\pi}{K}$
\State Set $\theta_n = {\rm arg max}_{\theta_{n}'\in \Phi_K} \cos(\theta_{n}' +\alpha_n - \phase{\mu})$, store $\theta_n$, $n=1,2,\ldots,N$
\State Set $g_0 = h_0 + \sum_{n=1}^N h_ne^{j\theta_n}$, ${\tt absgmax} = |g_0|$
\For{$l = 1, 2, \ldots, N$}
\State Let $(\theta_{n_l} + \omega \leftarrow \theta_{n_l}) \;\mathrm{mod}\, \Phi_K$
\State Let
\[
g_l = g_{l-1} + h_{n_l} \left(e^{j\theta_{n_l}} - e^{j(\theta_{n_l} - \omega)\, \mathrm{mod}\, \Phi_K}\right)
\]
\If{$|g_l| > {\tt absgmax}$}
\State Let ${\tt absgmax} = |g_l|$
\State Store updated $\theta_{n_l}$ 
\EndIf
\EndFor
\State Read out $\theta_n^*$ as the stored $\theta_n$, $n=1,2,\ldots,N$.
\end{algorithmic}
\end{algorithm}
\section{Fewer Than $N$ Steps Suffice When $|{\cal N}(\lambda_l)| > 1$ for Some $l$}
We have
\begin{equation}
s_{nk} = e^{j(\alpha_n+(k-\frac{1}{2})\frac{2\pi}{K})}
\end{equation}
where $\alpha_n\in[0,2\pi)$, $n=1,2,\ldots,N$, and $k=1,2,\ldots,K.$ Assume that there are repetitions among $s_{nk}$,
so that there are $M$ unique values of $s_{nk}$ with $0 \leq \lambda_1 < \cdots < \lambda_M < 2\pi$.
We define
\begin{equation}
{\cal N}(\lambda_l) = \{n | \lambda_l = \phase{s_{nk}}\}.
\end{equation}
Consider a repetition among $s_{nk}$, i.e., assume there are $n_1,$ $n_2,$ $k_1,$ and $k_2$ such that
$s_{n_1,k_1} = s_{n_2,k_2}$, i.e.,
\begin{equation}
\begin{aligned}
\left( \alpha_{n_1}+ \frac{(2k_1-1)\pi}{K}\right) & \;{\rm mod}\; 2\pi = \\
& \left( \alpha_{n_2}+\frac{(2k_2-1)\pi}{K}\right) \;{\rm mod}\; 2\pi .
\end{aligned}
\label{eqn:eqn44}
\end{equation}
Equation (\ref{eqn:eqn44}) is possible only if $\beta_{n_1}=\beta_{n_2}$ as $\beta_n$ are defined in Claim~1. Therefore,
all $K$ phase values represented by $\beta_{n_1}$ and $\beta_{n_2}$ must be equal, meaning there is an $N'$ such that
$M=(N-N')K$. Consequently, the problem of sorting $s_{nk}$ according to their phase values with $0\le \alpha_1 <
\alpha_2 < \cdots < \alpha_N < \frac{2\pi}{K}$ reduces to the following
\begin{equation}
s_{mk} = e^{j(\gamma_m+(k-\frac{1}{2})\frac{2\pi}{K})},\quad m=1, 2, \ldots ,\frac{M}{K},\; k = 1, 2, \ldots, K
\label{eqn:eqn47}
\end{equation}
where $\gamma_m = \min \{\alpha_{n_m,1},\alpha_{n_m,2},\ldots,\alpha_{n_m,G_m}\}$ such that $\beta_{n_m,1}=\beta_{n_m,2}
= \cdots = \beta_{n_m,G_m}$ and $0\le \gamma_1 < \gamma_2 < \cdots < \gamma_{\frac{M}{K}=N-N'} < \frac{2\pi}{K}$. So, this time,
there are $\frac{M}{K}+1 = N - N' + 1$ many cases.

For unique $s_{mk}$, let ${\cal M}(\lambda_l) = \{m|\lambda_l=\phase{s_{mk}}\}$. We know from (\ref{eqn:eqn387}) that
the following must hold
\begin{align}
\big({\cal M}(\lambda_l)\big) & _{l=1}^{M=(N-N')K} = \nonumber\\
\Big\{ & \underbrace{\frac{M}{K}-i+1,\ldots,\frac{M}{K},1, 2, \ldots, \frac{M}{K}-i}_{1},\nonumber\\
& \underbrace{\frac{M}{K}-i+1,\ldots,\frac{M}{K},1, 2, \ldots, \frac{M}{K}-i}_{2}, \\
& \hspace{26.5mm}\vdots \nonumber\\
& \underbrace{\frac{M}{K}-i+1,\ldots,\frac{M}{K},1, 2, \ldots, \frac{M}{K}-i}_{K}\Big\}\nonumber
\end{align}
for $i=1, 2 , \ldots, \frac{M}{K}$ where in each one of the $K$ groups there are $\frac{M}{K}=N-N'$ elements. To calculate
${\cal N}(\lambda_l)$, we define the following sets
\begin{equation*}
{\cal R}_m = \{ n_{m,1},n_{m,2},\ldots,n_{m,G_m} | \hspace{50mm}
\end{equation*}
\begin{equation}
\gamma_m = \min\{\alpha_{n_m,1},\alpha_{n_m,2},\ldots,\alpha_{n_m,G_m}\},\hspace{10mm}
\label{eqn:Rm}
\end{equation}
\begin{equation*}
\hspace{20mm}\beta_{n_m,1}=\beta_{n_m,2}=\cdots=\beta_{n_m,G_m}\}
\end{equation*}
where $G_m = |{\cal R}_m|$ and $\bigcup_{m=1}^\frac{M}{K}{\cal R}_m = \{1,2,\ldots,N\}$ must hold.
As a consequence, one can calculate
${\cal N}(\lambda_l) = {\cal R}_{{\cal M}(\lambda_l)}$. Therefore, the ``update loop'' in Algorithm~2
can be written as
\begin{align}
\big({\cal N}(\lambda_l)\big) & _{l=1}^{M=(N-N')K} = \nonumber\\
\Big\{ & \underbrace{{\cal R}_{\frac{M}{K}-i+1},\ldots,{\cal R}_\frac{M}{K},{\cal R}_1, {\cal R}_2, \ldots, {\cal R}_{\frac{M}{K}-i}}_{1},\nonumber\\
& \underbrace{{\cal R}_{\frac{M}{K}-i+1},\ldots,{\cal R}_\frac{M}{K},{\cal R}_1, {\cal R}_2, \ldots, {\cal R}_{\frac{M}{K}-i}}_{2},\label{eqn:eqn46}\\
& \hspace{26.5mm}\vdots \nonumber\\
& \underbrace{{\cal R}_{\frac{M}{K}-i+1},\ldots,{\cal R}_\frac{M}{K},{\cal R}_1, {\cal R}_2, \ldots, {\cal R}_{\frac{M}{K}-i}}_{K}\Big\}\nonumber
\end{align}
where the periodicity in the update rule still holds in (\ref{eqn:eqn46}), i.e., ${\cal N}(\lambda_{l'})={\cal N}(\lambda_{l'+\frac{M}{K}}),$ $l'=
1,2,\ldots,\frac{M}{K}(K-1)$. With the new update rule, after any $\frac{M}{K}$ consecutive steps in Algorithm~2, the phase shift selections will
be updated such that $\theta_n \rightarrow \theta_n + \omega,$ $n=1,2,\ldots,\frac{M}{K}=N-N'$.
This will result in
\begin{equation}
|g_{c,l'}| = \left|g_{c,l'+\frac{M}{K}}\right|, \quad l'=1,2,\ldots,\frac{M}{K}(K-1).
\label{eqn:eqn50}
\end{equation}
Therefore, the sufficiency of $\frac{M}{K}=N - N'$ steps follows from (\ref{eqn:eqn43}) and the text that follows it
in Section~\ref{sec:nsteps}.

Algorithm~\ref{alg:alg32} implements the technique described in this section. The initialization technique introduced in Appendix~\ref{sec:appSimpleUpdate}
for Algorithm~2 is employed in Algorithm~\ref{alg:alg32}.
\begin{algorithm}[!t]
\caption{Simplified Algorithm~2 where $|{\cal N}(\lambda_l)| > 1$ for some $l$}\label{alg:alg32}
\begin{algorithmic}[1]
\State {\bf Initialization:} Set $\phase{\mu}=\alpha_0 - \frac{\pi}{K}$ 
\State Find $\gamma_m$ and ${\cal R}_m$ as in (\ref{eqn:eqn47}) and (\ref{eqn:Rm}), $m=1,2,\ldots,\frac{M}{K}$
\State Compute $\varphi_m = \gamma_m - \alpha_0$ $(\mathrm{mod} \frac{2\pi}{K})$, $m=1,2,\ldots,\frac{M}{K}$ 
\State Sort $\varphi_m$ such that $0 \le \varphi_{m_1} < \varphi_{m_2} < \cdots < \varphi_{m_\frac{M}{K}} <\frac{2\pi}{K}$
\State Set $\theta_n = {\rm arg max}_{\theta_{n}'\in \Phi_K} \cos(\theta_{n}' +\alpha_n - \phase{\mu})$, store $\theta_n$, $n=1,2,\ldots,N$
\State Set $g_0 = h_0 + \sum_{n=1}^N h_ne^{j\theta_n}$, ${\tt absgmax} = |g_0|$
\For{$l = 1, 2, \ldots, \frac{M}{K}=N-N'$}
\State Let $(\theta_{n} + \omega \leftarrow \theta_{n}) \;\mathrm{mod}\, \Phi_K$ $n\in{\cal R}_{m_l}$
\State Let
\[
g_l = g_{l-1} + \sum_{n\in{\cal R}_{m_l}}h_{n}\left(e^{j\theta_{n}} - e^{j(\theta_{n} - \omega)\, \mathrm{mod}\, \Phi_K}\right)
\]
\If{$|g_l| > {\tt absgmax}$}
\State Let ${\tt absgmax} = |g_l|$
\State Store updated $\theta_{n}$ for $n\in {\cal R}_{m_l}$
\EndIf
\EndFor
\State Read out $\theta_n^*$ as the stored $\theta_n$, $n=1,2,\ldots,N$.
\end{algorithmic}
\end{algorithm}

Note that if the BS-UE link is completely blocked, i.e., $h_0=0$, the for loop in Step~7 can end at $l=\frac{M}{K}-1=N-N'-1$, which is one fewer step to run
Algorithm~4. This is because, we can guarantee in (\ref{eqn:eqn50}) that $|g_{c,l'}| = \left|g_{c,l'+\frac{M}{K}}\right|,$ whereas we cannot say right away that
$|g_{c,l'}+h_0| = \left|g_{c,l'+\frac{M}{K}}+h_0\right|$ will be satisfied.

A verification of this result is provided in Appendix~\ref{sec:appLessThanN}.

\section{Conclusion}
In this paper, we provided necessary and sufficient conditions for determination of optimum phase values in order to
maximize the received power at a UE which receives its transmission by means of reflections from an RIS, when the
phase values are from a discrete-valued set. Algorithms are provided to achieve this in a number of steps equal
to $N$, the number of RIS elements, or fewer. It is shown that the conditions that appeared in \cite{b1} do not
achieve the maximum received power. In addition, in \cite{b1}, the number of steps to achieve this maximum is given
as $KN$ or $2N$ on the average 
In conclusion, for a discrete-phase RIS, the techniques in this paper achieve the optimum received power in the
smallest number of steps published in the literature.
\vfill

\appendices
\section{Proof of Lemma~1}\label{sec:app-prooflemma}
In \cite{b1}, the proof of Lemma~1 reads as follows: ``Suppose that there exists
some
$\theta_n' \in \Phi_K \backslash \{\theta_n^*\}$ with
\begin{equation}
|(\theta_n' +
\alpha_n - \phase{\mu}) \;{\rm mod}\; 2\pi| < |(\theta_n^* + \alpha_n -
\phase{\mu}) \;{\rm mod}\; 2\pi|."
\label{eqn:c1}
\end{equation}
Note that $\theta_n^*$ is defined in
equation (11) of \cite{b1} as
\begin{equation}
\theta_n^* = {\rm arg}\ \underset{\theta_n\in
\Phi_K}\min |(\theta_n + \alpha_n -\phase{\mu}) \; {\rm mod}\; 2\pi |.
\label{eqn:c2}
\end{equation}
As a result, according to equation (11) of \cite{b1}
or ({\ref{eqn:c2}), it is not possible that a $\theta_n'$ as in
(\ref{eqn:c1}) exist.

A number of versions of \cite{b1} exist on arxiv.org: \cite{b11,b12,b13,b14,b15}. In
\cite{b11,b12,b13}, $\theta_n^*$ is given as in (\ref{eqn:c2}). In \cite{b14},
(\ref{eqn:c1}) is replaced with
\begin{equation}
\bigg| {\rm Arg}\bigg(\frac{h_ne^{j\theta_n'}}{h_0}\bigg)\bigg| < \bigg| {\rm Arg}\bigg(\frac{h_ne^{j\theta_n^*}}{h_0}\bigg)\bigg|,
\label{eqn:c5}
\end{equation}
and (\ref{eqn:c2}) is replaced with
\begin{equation}
\theta_n^* = {\rm arg}\ \underset{\theta_n\in \Phi_K}{\min} \bigg| {\rm Arg}\bigg(\frac{h_ne^{j\theta_n}}{h_0}\bigg)\bigg|.
\label{eqn:c4}
\end{equation}
On the other hand, in \cite{b15},
(\ref{eqn:c1}) is replaced with
\begin{equation}
\bigg| {\rm Arg}\bigg(\frac{h_ne^{j\theta_n'}}{\mu}\bigg)\bigg| < \bigg| {\rm Arg}\bigg(\frac{h_ne^{j\theta_n^*}}{\mu}\bigg)\bigg|,
\label{eqn:c7}
\end{equation}
and (\ref{eqn:c2}) is replaced with
\begin{equation}
\theta_n^* = {\rm arg}\ \underset{\theta_n\in \Phi_K}{\min} \bigg| {\rm Arg}\bigg(\frac{h_ne^{j\theta_n}}{\mu}\bigg)\bigg|.
\label{eqn:c6}
\end{equation}
Equation (\ref{eqn:c6}) is consistent with the development in \cite{PA23} as well as this newer version of \cite{PA23}. However,
a $\theta_n'$ as in (\ref{eqn:c7}) cannot exist because of the condition in (\ref{eqn:c6}). 
\section{Element-Based Simple Update Rule}\label{sec:appSimpleUpdate}
We now further simplify Algorithm~2, so that there is no need for calculating $s_{nk}$ or $\lambda_l .$
What we need to have is, given an initial $\mu_0=e^{j\phase{\mu_0}}$. say $\mu_0\in {\rm arc}
(e^{j\lambda_{i-1}}:e^{j\lambda_i}),$ we want to know the $N$-step update rule ${\cal N}(\lambda_l),
l=i,i+1,\ldots,i+N-1$ in the for loop of Algorithm~2.

{\em Claim:\/} Let ${\cal U}$ be the set to define the $N$ consecutive updates in the for loop of Algorithm~2.
For an initial $\phase{\mu_0},$ the update rule in the for loop of Algorithm~2 will be $\{{\cal U} = n_1,n_2,
\ldots, n_N | 0\le \varphi_{n_1} < \varphi_{n_2} < \cdots < \varphi_{n_N} < \frac{2\pi}{K}, \varphi_n =
(\alpha_n - \phase{\mu_0} + \frac{\pi}{K}) \;{\rm mod}\; \frac{2\pi}{K}, n = 1,2, \ldots, N\}$.

{\em Proof:\/} First, consider the case when $\phase{\mu_0} = 0.$ We know that the initial arc is
${\rm arc}(e^{j\lambda_L}:e^{j\lambda_1})$. Therefore, the update rule must be
${\cal U} = ({\cal N}(\lambda_l))_{l=1}^N$. We have already calculated this in (\ref{eqn:eqn387}) for any
Case $i$ given in (\ref{eqn:eqn34}). Note, from (\ref{eqn:eqn34}) to (\ref{eqn:eqn387}),
$({\cal N}(\lambda_l))_{l=1}^{NK}$ follows from the indexes of the
sorted values of
\begin{equation}
\varphi_n = \left(\alpha_n+\frac{\pi}{K}\right) \;{\rm mod}\; \frac{2\pi}{K} .
\label{eqn:appb1}
\end{equation}
Now, consider the case when $\phase{\mu_0}\notin {\rm arc}(e^{j\lambda_L}:e^j{\lambda_1})$. In this case,
instead of moving $\mu$ to a new arc, we can introduce an offset of $-\phase{\mu_0}$ for all $\lambda_l$.
Note that this corresponds to $\alpha_n\rightarrow \alpha_n - \phase{\mu_0},$ for all $n$. Therefore
(\ref{eqn:appb1}) will be updated as
\begin{equation}
\varphi_n = \left(\alpha_n - \phase{\mu_0} + \frac{\pi}{K}\right)\; {\rm mod} \frac{2\pi}{K}.
\label{eqn:appb2}
\end{equation}
Thus, the proof is complete.$\hfill\blacksquare$

Now, when we initialize $\phase{\mu} = \alpha_0 - \frac{\pi}{K}$, we can simply insert $\phase{\mu_0}
= \alpha_0 -\frac{\pi}{K}$ in (\ref{eqn:appb2}) and get
\begin{equation}
\varphi_n = (\alpha_n - \alpha_0) \;{\rm mod}\; \frac{2\pi}{K}
\end{equation}
to be used in the initialization step. It is important to note that, this simplification relieves Algorithm~2
from the burden to calculate $NK$ instances of both $s_{nk}$ and $\lambda_l$. Also, when the BS-UE link
is completely blocked, or $h_0=0$, initializations can be updated as $\phase{\mu}=0$ in Step~4 and
$\varphi_n = \left(\alpha_n-\frac{\pi}{K}\right)\; {\rm mod}\; \frac{2\pi}{K}$ for $n=1,2,\ldots,\frac{M}{K}$ in Step~2.

\section{Verification of $N-N'$ Steps When $|{\cal N}(\lambda_l)| \neq 1$}\label{sec:appLessThanN}
Assume there are replicas among $s_{nk}$. We know from (\ref{eqn:eqn44}) that there is a positive integer $N'$ such that
\begin{equation}
M = (N-N')K
\label{eqn:eqn48}
\end{equation}
where
\begin{eqnarray}
N' & = & \sum_{m=1}^{{M}/{K}}\left(|{\cal R}_m|-1\right)\\
 & = & \sum_{m=1}^{{M}/{K}} G_m - \frac{M}{K}
\end{eqnarray}
where $G_m = |{\cal R}_m|.$

Replacing $N'$ into (\ref{eqn:eqn48}), we get
\begin{equation}
\sum_{m=1}^{M/K}G_m = N.
\end{equation}
Since there are replicas among $s_{nk},$ there are $m$ for which $G_m>1$, from which $M/K<N$ follows as a necessary condition.

Hence, if $|{\cal N}(\lambda_l)| \neq 1,$ then the global optimum can be reached in $N-N'$ number of steps.


\bibliographystyle{IEEEtran}
\bibliography{ref}
\end{document}